\documentclass[12pt]{article}

\catcode`\@=11
\def\@citex[#1]#2{%
\if@filesw \immediate \write \@auxout {\string \citation {#2}}\fi
\@tempcntb\m@ne \let\@h@ld\relax \def\@citea{}%
\@cite{%
  \@for \@citeb:=#2\do {%
    \@ifundefined {b@\@citeb}%
      {\@h@ld\@citea\@tempcntb\m@ne{\bf ?}%
      \@warning {Citation `\@citeb ' on page \thepage \space undefined}}%
      {\@tempcnta\@tempcntb \advance\@tempcnta\@ne%
      \@tempcntb\number\csname b@\@citeb \endcsname \relax%
      \ifnum\@tempcnta=\@tempcntb 
        \ifx\@h@ld\relax%
          \edef \@h@ld{\@citea\csname b@\@citeb\endcsname}%
        \else%
          \edef\@h@ld{\ifmmode{-}\else--\fi\csname b@\@citeb\endcsname}%
        \fi%
      \else
        \@h@ld\@citea\csname b@\@citeb \endcsname%
        \let\@h@ld\relax%
      \fi}%
    \def\@citea{,\penalty\@highpenalty\,}%
  }\@h@ld
}{#1}}

\def\@citeb#1#2{{[#1]\if@tempswa , #2\fi}}
\def\@citeu#1#2{{$^{#1}$\if@tempswa , #2\fi }}
\def\@citep#1#2{{#1\if@tempswa , #2\fi}}

\def\bcites{         
        \catcode`\@=11
        \let\@cite=\@citeb
        \catcode`\@=12
}

\def\upcites{         
        \catcode`\@=11
        \let\@cite=\@citeu
        \catcode`\@=12
}

\def\plaincites{      
        \catcode`\@=11
        \let\@cite=\@citep
        \catcode`\@=12
}

\newcount\hour
\newcount\minute
\newtoks\amorpm
\hour=\time\divide\hour by 60
\minute=\time{\multiply\hour by 60 \global\advance\minute by-\hour}
\edef\standardtime{{\ifnum\hour<12 \global\amorpm={am}%
        \else\global\amorpm={pm}\advance\hour by-12 \fi
        \ifnum\hour=0 \hour=12 \fi
        \number\hour:\ifnum\minute<10 0\fi\number\minute\the\amorpm}}
\edef\militarytime{\number\hour:\ifnum\minute<10 0\fi\number\minute}

\def\draftlabel#1{{\@bsphack\if@filesw {\let\thepage\relax
   \xdef\@gtempa{\write\@auxout{\string
      \newlabel{#1}{{\@currentlabel}{\thepage}}}}}\@gtempa
   \if@nobreak \ifvmode\nobreak\fi\fi\fi\@esphack}
        \gdef\@eqnlabel{#1}}
\def\@eqnlabel{}
\def\@vacuum{}
\def\marginnote#1{}
\def\draftmarginnote#1{\marginpar{\raggedright\scriptsize\tt#1}}
\def\draft{
        \pagestyle{plain}
        \overfullrule=2pt
        \oddsidemargin -.5truein
        \def\@oddhead{\sl \phantom{\today\quad\militarytime} \hfil
        \smash{\Large\sl DRAFT} \hfil \today\quad\militarytime}
        \let\@evenhead\@oddhead
        \let\label=\draftlabel
        \let\marginnote=\draftmarginnote
        \def\ps@empty{\let\@mkboth\@gobbletwo
        \def\@oddfoot{\hfil \smash{\Large\sl DRAFT} \hfil}
        \let\@evenfoot\@oddhead}
        \def\@eqnnum{(\theequation)\rlap{\kern\marginparsep\tt\@eqnlabel}%
        \global\let\@eqnlabel\@vacuum}  }

\def\blackfonts{
        \font\blackboard=msbm10 scaled\magstep1
        \font\blackboards=msbm8
        \font\blackboardss=msbm6
}

\def\prep{         
        \catcode`\@=11
        \input art10.sty
        \catcode`\@=12
        
        \let\small\null
        \def\blackfonts{
                \font\blackboard=msbm10
                \font\blackboards=msbm7
                \font\blackboardss=msbm5
        }
        \let\sl\it
        \twocolumn
        \sloppy
        \voffset=-2.54truecm
        \hoffset=-2.54truecm
        \flushbottom
        \parindent 1em
        \leftmargini 2em
        \leftmarginv .5em
        \leftmarginvi .5em
        \marginparwidth 48pt
        \marginparsep 10pt
        \setlength{\columnsep}{2truecm}
        \setlength{\textwidth}{25.4truecm}
        \setlength{\textheight}{17truecm}
        \baselineskip=16pt
        \oddsidemargin .18truein
        \evensidemargin .17truein
}

\def\eqalign#1{\null\,\vcenter{\openup\jot\m@th
  \ialign{\strut\hfil$\displaystyle{##}$&$\displaystyle{{}##}$\hfil
      \crcr#1\crcr}}\,}
\def\eqalignno#1{\displ@y \tabskip\centering
  \halign to\displaywidth{\hfil$\@lign\displaystyle{##}$\tabskip\z@skip
    &$\@lign\displaystyle{{}##}$\hfil\tabskip\centering
    &\llap{$\@lign##$}\tabskip\z@skip\crcr
    #1\crcr}}

\def\section{\@startsection {section}{1}{\z@}{3.ex plus 1ex minus
 .2ex}{2.ex plus .2ex}{\large\bf}}
\def\subsection{\@startsection{subsection}{2}{\z@}{2.75ex plus 1ex minus
 .2ex}{1.5ex plus .2ex}{\bf}}        

\def\appendix{{\newpage\section*{Appendix}}\let\appendix\section%
        {\setcounter{section}{0}
        \gdef\thesection{\Alph{section}}}\section}

\def\abstract{\if@twocolumn
\section*{Abstract}
\else 
\begin{center}
{\bf Abstract\vspace{-.5em}\vspace{0pt}}
\end{center}
\quotation
\fi}

\catcode`\@=12

\def\sqr#1#2{{\vcenter{\vbox{\hrule height.#2pt\hbox{\vrule width.#2pt 
height#1pt \kern#1pt \vrule width.#2pt}\hrule height.#2pt}}}}

\def\=d{\,{\buildrel\rm def\over =}\,}

\def\i3p{\p32\int d^3p}

\def\As{A\hbox to 1pt{\hss /}}
\def\np4{\int d^4p_1\cdots d^4p_{n-1}\, }

\def\nx4{\int d^4x_1\ldots d^4x_n\, }

\def\kon#1#2{\vbox{\halign{##&&##\cr
\lower4pt\hbox{$\scriptscriptstyle\vert$}\hrulefill &
\hrulefill\lower4pt\hbox{$\scriptscriptstyle\vert$}\cr $#1$&
$#2$\cr}}}

\def\konv#1#2#3{\hbox{\vrule height12pt depth-1pt}
\vbox{\hrule height12pt width#1cm depth-11.6pt}
\hbox{\vrule height6.5pt depth-0.5pt}
\vbox{\hrule height11pt width#2cm depth-10.6pt\kern5pt
      \hrule height6.5pt width#2cm depth-6.1pt}
\hbox{\vrule height12pt depth-1pt}
\vbox{\hrule height6.5pt width#3cm depth-6.1pt}
\hbox{\vrule height6.5pt depth-0.5pt}}
\def\konu#1#2#3{\hbox{\vrule height12pt depth-1pt}
\vbox{\hrule height1pt width#1cm depth-0.6pt}
\hbox{\vrule height12pt depth-6.5pt}
\vbox{\hrule height6pt width#2cm depth-5.6pt\kern5pt
      \hrule height1pt width#2cm depth-0.6pt}
\hbox{\vrule height12pt depth-6.5pt}
\vbox{\hrule height1pt width#3cm depth-0.6pt}
\hbox{\vrule height12pt depth-1pt}}

\def\konw#1#2#3{\hbox{\vrule height12pt depth-1pt}
\vbox{\hrule height12pt width#1cm depth-11.6pt}
\hbox{\vrule height6.5pt depth-0.5pt}
\vbox{\hrule height12pt width#2cm depth-11.6pt \kern5pt
      \hrule height6.5pt width#2cm depth-6.1pt}
\hbox{\vrule height6.5pt depth-0.5pt}
\vbox{\hrule height12pt width#3cm depth-11.6pt}
\hbox{\vrule height12pt depth-1pt}}

\def\i{{\rm int}}

\def\m3{{\mu_1\mu_2\mu_3}}

\def\p{{(+)}}

\def\be{\begin{equation}}       \def\eq{\begin{equation}}
\def\ee{\end{equation}}         \def\eqe{\end{equation}}

\def\bea{\begin{eqnarray}}      \def\eqa{\begin{eqnarray}}
\def\ena{\end{eqnarray}}        \def\eea{\end{eqnarray}}
                                \def\eqae{\end{eqnarray}}

\def\ba{\begin{array}}
\def\ea{\end{array}}
\def\unit{1 \hskip-.3em \raise2pt\hbox{$ \scriptstyle |$ } }

\def\i{\iota}

\def\m{\mu}

\def\p{\pi}                
\def\t{\tau}

\def\bop#1{\setbox0=\hbox{$#1M$}\mkern1.5mu
        \vbox{\hrule height0pt depth.04\ht0
        \hbox{\vrule width.04\ht0 height.9\ht0 \kern.9\ht0
        \vrule width.04\ht0}\hrule height.04\ht0}\mkern1.5mu}

\def\>{\rangle} 

\def\<{\langle} 
\def\Dsl{D \hskip-.6em \raise1pt\hbox{$ / $ } }

\def\sl#1{\rlap{\hbox{$\mskip 1 mu /$}}#1}
\def\leftrightarrowfill{$\mathsurround=0pt \mathord\leftarrow \mkern-6mu
       \cleaders\hbox{$\mkern-2mu \mathord- \mkern-2mu$}\hfill
       \mkern-6mu \mathord\rightarrow$}
\def\dvec#1{\vbox{\ialign{##\crcr
       \leftrightarrowfill\crcr\noalign{\kern-1pt\nointerlineskip}
       $\hfil\displaystyle{#1}\hfil$\crcr}}}          
\def\hook#1{{\vrule height#1pt width0.4pt depth0pt}}
\def\leftrighthookfill#1{$\mathsurround=0pt \mathord\hook#1
       \hrulefill\mathord\hook#1$}
\def\underhook#1{\vtop{\ialign{##\crcr                 
       $\hfil\displaystyle{#1}\hfil$\crcr
       \noalign{\kern-1pt\nointerlineskip\vskip2pt}
       \leftrighthookfill5\crcr}}}
\def\smallunderhook#1{\vtop{\ialign{##\crcr      
       $\hfil\scriptstyle{#1}\hfil$\crcr
       \noalign{\kern-1pt\nointerlineskip\vskip2pt}
       \leftrighthookfill3\crcr}}}

\def\sfrac#1#2{{\vphantom1\smash{\lower.5ex\hbox{\small$#1$}}\over
       \vphantom1\smash{\raise.4ex\hbox{\small$#2$}}}} 
\def\bfrac#1#2{{\vphantom1\smash{\lower.5ex\hbox{$#1$}}\over
       \vphantom1\smash{\raise.3ex\hbox{$#2$}}}}      
\def\afrac#1#2{{\vphantom1\smash{\lower.5ex\hbox{$#1$}}\over#2}}  
\def\on#1#2{{\buildrel{\mkern2.5mu#1\mkern-2.5mu}\over{#2}}}
\def\ddt#1{\on{\hbox{\LARGE .\kern-2pt.}}#1}             
\def\tdt#1{\on{\hbox{\LARGE .\kern-2pt.\kern-2pt.}}#1}   

\def\boxes#1{
       \newcount\num
       \num=1
       \newdimen\downsy
       \downsy=-1.5ex
       \mskip-2.8mu
       \bo
       \loop
       \ifnum\num<#1
       \llap{\raise\num\downsy\hbox{$\bo$}}
       \advance\num by1
       \repeat}
\def\boxup#1#2{\newcount\numup
       \numup=#1
       \advance\numup by-1
       \newdimen\upsy
       \upsy=.75ex
       \mskip2.8mu
       \raise\numup\upsy\hbox{$#2$}}

\newskip\humongous \humongous=0pt plus 1000pt minus 1000pt
\def\caja{\mathsurround=0pt}
\def\eqalign#1{\,\vcenter{\openup2\jot \caja
       \ialign{\strut \hfil$\displaystyle{##}$&$
       \displaystyle{{}##}$\hfil\crcr#1\crcr}}\,}
\newif\ifdtup

\def\1ov4{{1\over 4}}

\def\ddt{\dot{\t}}

\newcommand{\rmd}{{\rm d}}

\newcommand{\beq}{\begin{equation}}
\newcommand{\eeq}{\end{equation}}

\def\ba{\begin{eqnarray}}
\def\ea{\end{eqnarray}}

\begin{document}

\null\vskip-50pt
\hfill KL-TH 00/05 
\vskip-10pt
\hfill UTHET-00-0701
\vskip-10pt
\hfill {\tt hep-th/0008198}
\vskip0.3truecm
\begin{center}
\vskip 1truecm
{\Large\bf
Lee-Yang edge singularity in the three-dimensional Gross-Neveu model
at finite temperature 
\\}
\vskip 1.2truecm
{\large {\bf Anastasios C. Petkou}$^{\dagger}$ \footnote{{\it e-mail}:
 petkou@physik.uni-kl.de} and
 {\bf George Siopsis}}$^{\star}$ \footnote{{\it e-mail}:
 gsiopsis@utk.edu} \\
\vskip 1truecm
$^{\dagger}$ Department of Theoretical Physics, University of
 Kaiserslautern\\ 
Postfach 3049, 67663 Kaiserslautern, Germany;  \\ 
$^{\star}$ Department  of
Physics and Astronomy, The University of Tennessee\\ Knoxville, TN
 37996-1200 U.S.A.
\end{center}
\vspace{1.2cm}

\noindent{\centerline{ \bf Abstract}}

We discuss the relevance of the Lee-Yang edge singularity to the
finite-temperature $Z_2$-symmetry
restoration transition of the Gross-Neveu model in three
dimensions. We present an explicit result for
its large-$N$ free-energy 
density in terms of $\zeta(3)$ and the absolute maximum of Clausen's
function. 

\vspace{1.5cm}

The Gross-Neveu model in $d=3$ dimensions provides a remarkable
example of a second order temperature driven phase transition in  a
theory which also exhibits dynamical symmetry breaking. The latter
property is purely
quantum field theoretical while the former one
involves classical thermal fluctuations.
We consider here the standard Lagrangian describing the Euclidean
version of the Gross-Neveu model with $U(2N)$ symmetry
\cite{zinn-justin,rosenstein,tassos1} 
\beq
\label{lagr}
{\cal L} = \bar{\psi}^a\slash\!\!\!\partial\psi^a
  +\frac{G_0}{2N}(\bar{\psi}^a\psi^a)^2\,,\,\,\,\,\,\,\,\, a=1,2,..,N\,, 
\eeq
where $G_0$ is the coupling.
The partition function for the theory can be written, after
integrating out the fundamental four-component 
massless Dirac fermions $\bar{\psi}^a\,,\,\psi^a$, 
with the help of the auxiliary scalar field $\sigma(x)$ as
\cite{zinn-justin,rosenstein,tassos1} 
\beq
\label{pf}
Z_{\sigma}[G_0]  =   \int ({\cal D}\sigma)\,e^{N 
\left[ 2\,{\rm Tr}[\ln  
    (-\partial^2 +\sigma^2)] -\frac{1}{2G_0}\int\rmd^3
    x\,\sigma^2(x)\right]}.
\eeq
The model possesses a discrete $Z_2$
``chiral'' symmetry as (\ref{lagr}) is invariant under
$\psi\rightarrow \gamma_5\psi$. The usual $1/N$ expansion is generated
if one expands as
$\sigma(x)=\sigma_0+O(1/\sqrt{N})$, provided that $\sigma_0$ satisfies the gap
equation
\beq
\label{gap0}
\frac{\sigma_0}{G_0}=\int\frac{\rmd^3
  p}{(2\pi)^3}\frac{4\sigma_0}{p^2+\sigma_0^2}\,.
\eeq
One renormalizes (\ref{gap0}) by introducing an UV cut-off $\Lambda$ as
\beq
\label{renorm}
\frac{1}{G_0}=\frac{1}{G_*}+\frac{1}{G_R} = 4\int^{\Lambda}\!\!\frac{\rmd^3
p}{(2\pi)^3}\frac{1}{p^2} +\frac{1}{G_R}\,, 
\eeq
and obtains the renormalized coupling
$1/G_R = -M/\pi$, where $M$ is 
the arbitrary mass scale introduced   
by renormalization. From (\ref{pf}) and (\ref{renorm}) it is easy
  to calculate the 
leading-$N$ renormalized ``effective action'' $V_R(\sigma_0;G_R)$
defined as
\beq
\label{effa}
\int\rmd^3 x \,V_R(\sigma_0;G_R) =
-\ln\left[\frac{Z_{\sigma_0}[G_R]}{Z_0[0]}\right]\,,
\eeq
where the subtraction on the r.h.s. of (\ref{effa}) ensures a finite
result. The result is  
\cite{rosenstein} 
\beq
\label{effa0}
V_R(\sigma_0,G_R) = \frac{N}{2\pi}\left(
  \frac{2}{3}|\sigma_0|^3 -M\sigma_0^2\right) \,.
\eeq
From (\ref{effa0}) we can clearly separate three regimes: {\bf A)}
For $M<0$ the minimum of (\ref{effa0}) is always at the origin and
  the theory is in the $Z_2$-{\it symmetric phase} with $\sigma_0=0$. {\bf B)}
For $M>0$ the minimum of (\ref{effa0}) is at $\sigma_0=M$, the theory is in the
$Z_2$-{\it broken phase} and $M$ can be identified as the mass of the
elementary fermionic fields. {\bf C)} Finally, for $M=0$, the theory
is at the {\it critical point} and it is a non-trivial
three-dimensional conformal field theory
(CFT). 

Notice that the UV subtraction prescription in (\ref{effa}) has
  generated the $|\sigma_0|^3$ term in (\ref{effa0}). 
  This term manifests itself as the dominant contribution at the
  critical point $M=0$. It is then conceivable that, to leading-$N$, the
  critical behavior of the model is somehow related to a
  $\phi^3$ theory.  Of course, the true critical ground
  state is at $\sigma_0=0$ which would correspond to the
  zero-coupling critical point of the  $\phi^3$ theory,
  or equivalently to a free-field theory. This is consistent with the
  well known mean
  field theory behavior, to
  leading-$N$, of all the critical quantities of the model 
  \cite{rosenstein}.

On the other hand, it is well known \cite{fisher,cardy} that the IR
limit of the theory
with action 
\beq
\label{phi3}
S=-\int[\frac{1}{2}(\partial\phi)^2+{\rm
i}(h-h_c)\phi+\frac{1}{3!}\lambda\phi^3]\,\rmd^d x\,, 
\eeq
 dictates the critical behavior
of an Ising model in a purely imaginary magnetic field
${\rm i}h_c$ as the critical temperature is approached from above
(from the symmetric phase). This critical point (Lee-Yang edge
singularity \cite{leeyang}) corresponds to a non-unitary theory  as
it involves an imaginary coupling constant $\lambda$. 

If the critical behavior of the Gross-Neveu
model is in any way related to a  $\phi^3$ theory, one would expect
that the Lee-Yang singularity might become relevant as one approaches the
critical point of the model from a suitable symmetric phase.
To investigate such a possibility we introduce the ingredient of
temperature $T$ by putting the model (\ref{pf}) in a slab geometry with
one finite dimension of 
length $L=1/T$. A
crucial point is that the 
renormalization (\ref{renorm}) is unaffected, since renormalizing the
theory in the 
bulk suffices to remove the UV-divergences for finite temperature
\cite{zinn-justin}. However, the gap equation (\ref{gap0}) now becomes
\ba
\label{gapt}
\frac{\sigma_0}{G_0} & = & 4\frac{\sigma_0}{L}
\sum_{n=0}^{\infty} \int^{\Lambda}\!\!\frac{\rmd^2 p}{(2\pi)^2}
\frac{1}{p^2+\omega_n^2 +\sigma_0^2} \nonumber\\
 & = & \frac{\sigma_0}{G_*} -\frac{\sigma^2_0}{\pi}-
 \frac{2\sigma_0}{\pi
   L}\ln\left(1+e^{-L\sigma_0}\right) \,.
\ea
Then, from (\ref{pf}) and (\ref{gapt}) we can
explicitly calculate the 
leading-$N$ renormalized ``effective action'' $V_R(\sigma_0,L;G_R)$ -
now depending in addition on the ``inverse'' temperature $L$ - as 
\ba
\label{effat}
V_R(\sigma_0,L;G_R) & =& \frac{N}{2\pi L^3} \Biggl
   [ \frac{2}{3}\sigma_0^3L^3-M\sigma_0^2L^3
   \nonumber \\
& & \hspace{-2cm}
   +4Li_3\left(-e^{-L\sigma_0}\right)-4\ln\left(e^{-L\sigma_0}\right) 
    Li_2\left(-e^{-L\sigma_0}\right) 
    \Biggl]
\ea
where $Li_n(z)$ are the standard polylogarithms \cite{lewin}. This
   ``effective action'' 
   presents a remarkably explicit example of high-temperature symmetry
   restoration in a {\it quantum} field theoretic
   system\footnote{Notice that although we are dealing with symmetry 
     restoration in 
     two dimension, the Mermin-Wanger-Coleman theorem is not violated
     as the relevant symmetry is discrete ($Z_2$ here).} which is
   {\it ordered} ($M>0$) at $T=0$.  The critical temperature is
   $1/L_c=T_c=M/2\ln 2$ 
   \cite{rosenstein}. 

When $M=0$ in (\ref{effat}), then for all
   $T>0$ we approach the critical point from the symmetric
   phase. This is the regime where we would expect the appearance of
   the Lee-Yang singularity. This seems rather
   difficult to imagine as, despite  the appearance in (\ref{effat})
   of the cubic term  
   $\sigma_0^3$ as a result of the UV subtraction prescription
   (\ref{effa}), the  coefficient of
   this  term is real. 
    Nevertheless, one can show that (\ref{effat}) for $M=0$ is in fact  an {\it
      even} function of $\sigma_0$.  To see this we express
    (\ref{effat}) in terms of  
   Nielsen's generalized polylogarithms $S_{n,p}(z)$
   \cite{ogreid} as follows 
\ba
\label{nielsen}
V_R(\sigma_0,L;0) & = & \frac{2N}{\pi L^3}\left
[ S_{1,2}(z)+S_{1,2}\left(\frac{1}{z}\right) -\zeta(3)\right]\\
S_{1,2}(z) & = & \frac{1}{2}\int_0^z \!\frac{\ln^2(1-y)}{y}\,\rmd y\\
S_{1,2}(1) &=&8\,S_{1,2}(-1)\,=\,\zeta(3)
\ea
where we have set $z=-e^{-L\sigma_0}$. From (\ref{nielsen}) we see
that $V_R(-\sigma_0,L;0) = V_R(\sigma_0,L;0)$. This remarkable
property 
   means that although the $L\rightarrow \infty$ ($T\rightarrow 0$)
   behavior of (\ref{effat}) looks like its is dominated by the cubic
   term (with real coefficient and 
ground state at 
   $\sigma_0=0$), fluctuations become important for {\it all} $T>0$
   and completely change the relevant underlying effective
   potential. To this end we point out that the step from
   (\ref{effat}) to (\ref{nielsen}) involves an all order resummation
   in $\sigma_0$. drive the theory towards another critical point. From
   (\ref{nielsen}). Then, from (\ref{nielsen}) we conclude  that away
   from $\sigma_0=0$ the 
   critical theory is described by
   an effective Hamiltonian which is an {\it even} function of
   $\sigma_0$. If we view now $\sigma_0$ as a {\it scalar order
     parameter} and couple it to an {\it external magnetic field}, the
   critical behavior of such a system in the high temperature phase
   can be shown  
   to correspond to a $\phi^3$ theory with purely imaginary coupling
   \cite{fisher}. The critical point is determined by the
   non-zero solution of the gap equation (\ref{gapt}) as
\beq
\label{eos}
\sigma_0 \left[\sigma_0+\frac{2}{L}\ln\left(1+e^{-\sigma_0L}\right)
\right]=0\Rightarrow \sigma_0=\pm{\rm i}\frac{2\pi}{3L}\,,
\eeq
where we restricted $-{\rm i}\pi < L\sigma_0 < {\rm i}\pi$ to avoid
the cut of the logarithm. The fact that $\sigma_0$ is now purely
imaginary, however, does not affect the reality properties of the
effective potential and we obtain 
\beq
\label{claus}
V_R(\pm{\rm i}\frac{2\pi}{3L},L;0) = \frac{N}{2\pi
L^3}\left[\frac{4}{3}\zeta(3)-\frac{8\pi}{3}Cl_2\left( \frac{\pi}{3}\right)
\right] \,,
\eeq
where $Cl_2(\theta)={\rm Im}\left[Li_2(e^{{\rm i}\theta})\right]$ is
Clausen's function \cite{lewin}. It is amusing to point out that
$Cl_2(\pi/3)\approx 1.014942..$ is the absolute maximum of Clausen's
function which is a   
well-documented numerical constant. 

Our result (\ref{claus}) corresponds to the leading-$N$ free-energy
density of the Lee-Yang 
edge singularity in $d=3$. The parameter $N$ should not be confused
with the number of components of the underlying order parameter
\cite{itzdrouffe}, but should be regarded as a suitable expansion
parameter such that (\ref{claus}) is the leading approximation to the
exact value of the free-energy density. Moreover, our result
(\ref{claus}) corresponds to
a new CFT in three-dimensions. 
Indeed, on general grounds \cite{cardy2,appelquist} one expects that
the free-energy density of a CFT placed in a slab geometry with one
finite dimension of length $L$ behaves as
$f_L-f_{\infty}=-\tilde{c}\,\Gamma(d/2) \zeta(d)/\pi^{d/2}L^d$. In
$d=2$ the
parameter $\tilde{c}$ is proportional to the central charge and the
conformal anomaly \cite{cardy3}. However, corresponding results in
$d>2$ are still unknown. In $d=3$ one easily obtains $\tilde{c}=3N$ for the
case of $N$ free massless four-component Dirac fermions
\cite{tassos1}. The value of $\tilde{c}$ for the Lee-Yang edge
singularity which can be read-off from (\ref{claus}) is larger than
$3N$, implying  that the corresponding CFT is non-unitary. This is  in
accordance with the two-dimensional results \cite{cardy}. 

In may cause
some worry that we have connected the critical behavior of a unitary
theory (Gross-Neveu) with a non-unitary one. Nevertheless, this is not
a direct connection. Staying within the Gross-Neveu model and starting
e.g. from the low temperature {\it broken} phase, we do not expect the
appearance of the Lee-Yang critical behavior studied
above. Namely, as we raise the temperature we simply expect that the
$Z_2$ symmetry is restored at the critical temperature $T_c=M/2\ln 2$
and then the 
system continues to be in the high-temperature symmetric temperature
phase for  all $T > T_c$. However, if we consider the Gross-Neveu model
as a component of some enlarged theory, it is quite conceivable that the
presence of other fields (e.g. gauge fields), or chemical potentials
might account for a possible Lee-Yang critical behavior 
at $T>T_c$ as they could induce imaginary values for the minimum of
the effective potential $\sigma_0$ \cite{tassos2}. Clearly,
the enlarged system should still be described by 
a unitary theory. From this point of view, we expect our approach
and results to be most suitable for discussing effects such as
the recently studied symmetry nonrestoration \cite{kogut}, since the
latter is related to an imaginary 
chemical potential. Our
leading-$N$ calculations reproduce the well-known 
mean field theory results for the Lee-Yang edge singularity critical
exponents. It would 
then be interesting to extend our results to next-to-leading order in $1/N$
for comparison with existing numerical calculations \cite{fisher2}.  

\section*{Acknowledgments}
A. C. P. is supported by Alexander von Humboldt Foundation and G. S. by
the US Department of Energy under grant
DE-FG05-91ER40627. A. C. P. would like to thank the University of
Tennessee, Knoxville for
its kind hospitality.

\end{document}